# Conductance modulation in topological insulator $Bi_2Se_3$ thin films with ionic liquid gating

Jaesung Son,[1] Karan Banerjee,[1] Matthew Brahlek,[2] Nikesh Koirala,[2] Seoung-Ki Lee,[3,4] Jong-Hyun Ahn,[4] Seongshik Oh,[2] and Hyunsoo Yang[1,a)]

[1]*Department of Electrical and Computer Engineering, National University of Singapore, 4 Engineering Drive 3, Singapore 117576, Singapore*

[2]*Department of Physics & Astronomy, Rutgers, The State University of New Jersey, 136 Frelinghuysen Road, Piscataway, New Jersey 08854, USA*

[3]*School of Advanced Materials Science and Engineering, Sungkyunkwan University, 300, Suwon, 440-746, Korea*

[4]*School of Electrical & Electronic Engineering, Yonsei University, Seoul 120-749, Korea*

A $Bi_2Se_3$ topological insulator field effect transistor is investigated by using ionic liquid as an electric double layer gating material, leading to a conductance modulation of 365% at room temperature. We discuss the role of charged impurities on the transport properties. The conductance modulation with gate bias is due to a change in the carrier concentration, whereas the temperature dependent conductance change is originated from a change in mobility. Large conductance modulation at room temperature along with the transparent optical properties makes topological insulators as an interesting (opto)electronic material.

[a)] e-mail address: eleyang@nus.edu.sg

Three-dimensional (3D) topological insulators (TIs) are a new state of quantum matter with conducting surfaces and an energy bandgap in the bulk. The unique physical surface properties, including a single Dirac cone at low energies and massless Dirac fermions, were reported from various 3D TI materials such as $Bi_2Se_3$, $Bi_2Sb_3$, $Bi_2Te_3$, and $Bi_{1.5}Sb_{0.5}Te_{1.7}Se_{1.3}$.[1-5] The surface state is protected from backscattering by time reversal symmetry, and spin momentum-locking is expected to provide highly spin polarized electron transport.[4,6] These characteristics make TIs as a promising candidate for future electronics and spintronics applications with low power consumption. Among the TIs discovered so far, $Bi_2Se_3$ is considered as one of the most promising TI materials with its well defined single Dirac cone and a larger bulk bandgap of 0.3 eV compared to that of 0.14 eV in $Bi_2Te_3$.[7] Moreover, considering that prototypes of a field effect transistor (FET) device have been fabricated with graphene by taking advantage of the massless behaviors of carriers,[8,9] TIs can be another suitable material for FET applications.

In recent studies of TI FET devices, large tunability of the Fermi level has been reported with a back gate, investigating the transport near the surface Dirac point at *low* temperatures.[10,11] The carrier type can be changed from n-type to p-type with the gate voltage.[12] However, the conductance modulation ratio at *room* temperature remains small. Recent studies show that 76% of conductance was modulated in $Bi_2Te_3$ at room temperature by using simultaneous dual gate control[13] and even smaller modulation was reported at ~200 K in $Bi_2Se_3$ film.[14] The observation of low conductance modulation is mainly due to a significant conductance through the heavily doped bulk originated from structural defects of TIs, such as vacancies or impurities.[10] Meanwhile, the advances in growing ultrathin TI films using molecular beam epitaxy (MBE) have shown dominant surface transport properties.[15] Moreover, electric double layer gating method using ionic liquid is a promising technique to overcome the limit of conventional high-κ dielectric due to its intrinsic high capacitance.[16] The ionic liquid exhibits an extremely huge capacitance, typically on the order of 1 – 20 µF/cm$^2$, and hence large conductance modulation can be observed in the FET channel due to the effective control of a carrier density with the ionic liquid gate dielectric.[17,18] In this letter, we demonstrate a large conductance modulation ratio from high quality Bi$_2$Se$_3$ thin films grown by MBE on sapphire substrates. Using ionic liquids as gate dielectrics, the highest modulation ratio of ~365% is achieved with a gate voltage of −1 V at room temperature. In addition, we identify the role of inhomogeneous charged impurities on the transport properties.

Large-area, high-quality Bi$_2$Se$_3$ films were grown on 3 inch c-axis Al$_2$O$_3$ substrates with MBE using the recently developed two-step scheme.[15] Optical transmission of as-grown TI films with different thicknesses was characterized using a spectrophotometer. Hall bar patterns were fabricated by optical lithography followed by Ar ion milling. Six electrodes were deposited by sputtering Ta/Cu (4 nm/80 nm) on top of the Hall bar pattern. A patterned ionic top-gate dielectric was prepared using a photolithographic technique.[19] The UV-cross-linkable precursor solution was prepared by mixing 1-ethyl-3-methylimidazolium bis(tri-fluoromethylsulfonyl)imide ([EMIM][TFSI]) ionic liquid, poly(ethyleneglycol) diacrylate (PEG-DA), and 2-hydroxy-2-methylpropiophenone (HOMPP) with a ratio of 88:8:4 (w/w) and dropped on a substrate. The radiation of UV results in chemically cross-linked solid electrolytes termed "ion gel" due to polymerization among acrylate end groups on PEG. The unexposed area is washed by deionized water.[19]

In analogy to the optical properties of graphene, a thin TI film has been predicted to be transparent and conductive for optoelectronic applications such as displays, organic light-emitting diodes (OLEDs), and touch screens.[8,20] The transparency of thin Bi$_2$Se$_3$ films was investigated using ultraviolet-visible-infrared (UV-VIS-IR) spectroscopy. Figure 1(a) shows that the net transmittance of the $Bi_2Se_3$ films decreases as the film becomes thicker. In order to extract the pure optical response from the TI films, transmittance is normalized with respect to the sapphire substrate. It is known that the other transparent conductive oxides such as indium tin oxide (ITO) or fluorinated tin oxide (FTO) become optically opaque in the IR regime due to free-electron plasma resonance.[21] However, most noticeably, both 5 quintuple layer (QL; 1 QL ≈ 1 nm) and 10 QL $Bi_2Se_3$ films show a relatively constant transmission of ~75% and 65%, respectively, in the IR range. It was reported that a 6 nm thick $Bi_2Se_3$ film grown in a furnace (Bridgeman method) exhibited 80% of optical transmission in 1 − 3 μm wavelength, which is in line with our observation, due to forbidden photoexcitation of carriers within the surface state of a TI originated from the non-spin degeneracy.[20] Interestingly, transmission drop at ~600 nm is observed which is similar to the previous study.[20] As the corresponding wavelength to the bulk bandgap of $Bi_2Se_3$ (0.3 eV) is ~4 μm, the transmission drop at 600 nm is unlikely related to photoabsorption at the band edges. Future studies are required to understand a transmission dip at 600 nm. For example, typical surface plasmon excitation in gold nanoparticles occurs at a resonance wavelength near 600 nm.[22] Nonetheless, the high transparency of thin TI films in a wide range of wavelength makes it as a possible candidate as not only transparent conductive materials but also transparent transistors.

Electrical measurements have been carried out in the $Bi_2Se_3$ films. The resistance ($R_{xx}$) as a function of temperature is plotted from a Hall bar device of 20 QL films as shown in Fig. 1(b). The resistance decreases as temperature decreases from 300 to 30 K. Below 30 K the resistance remains almost constant as reported previously.[15] Here, the temperature dependent resistance indicates metallic transport, since the TI channel is n-doped without gate electric fields.[11] The Hall measurement data ($R_{xy}$) at 2 K are shown in Fig. 1(c). The carrier concentration ($n_{2D}$) and the type of carrier have been calculated from the magnitude and sign of the slope of $R_{xy}$(H), respectively, using $n_{2D} = 1/eR_{xy}$, where $R_{xy}$ is the cross channel resistance with a current of 20 µA, $e$ is the electron charge, and $H$ is the applied magnetic field normal to the sample plane. The observed linear $R_{xy}$(H) curve indicates that one type of carriers is dominant in our device, which is n-type with a carrier concentration of 4.23 × 10$^{13}$/cm$^2$. The mobility was estimated to be 544 cm$^2$/V·s. In the previous studies, the effective electron mobility of a exfoliated Bi$_2$Te$_3$ single crystal flake was estimated to be ~17 cm$^2$/V·s,[13] while 14 nm thick Bi$_2$Se$_3$ single crystal flakes show a mobility of 300 – 1200 cm$^2$/V·s.[23] The estimated carrier mobility of our Bi$_2$Se$_3$ films is in a similar range with the earlier reports.[15] Figure 1(d) shows the voltage ($V_{xx}$) of the channel as a function of the source-drain current ($I_{sd}$) without gate bias at 2 K and 300 K. Note that there is no hysteresis in the I-V curves, whereas we observe the hysteretic I-V data in electrolyte-gated devices as discussed later.

Applying ion gel as a gate dielectric as shown in Fig. 2(a) allows an efficient control of the carrier density with low operating voltages. A coplanar-gate configuration is used to apply gate voltage.[24] Figure 2(b) shows the resistance modulation with gate bias at room temperature in 20 QL devices. The resistance increases as the gate voltage decreases and it saturates below -1 V. The leakage current through the ion gel dielectric is less than 150 nA in the voltage scanning ranges. The observed hysteresis can be attributed to slow ion motion (slow polarization relaxation of gel dielectrics) and the interfacial trap states.[19] The transfer characteristics are reproducible, as the curves of the first scan up to -0.8 V is very similar to the fourth scan with the same voltage range after applying higher voltage sweeps. The conductance modulation ratio in Fig. 2(b) increases with increasing the sweeping ranges; a 170% of modulation with a gate bias of -0.8 V and ~365% with a gate bias of -1 – -1.5 V, which is much larger values than other studies reported at *room* temperature. For example, ~2% and ~76% of resistance modulation have been reported in $Bi_2Se_3$ and $Bi_2Te_3$ FET respectively.[13,14]

In order to estimate the gate voltage corresponding to the Dirac point, we use the relation, $n_{2D} = -C_g(V_g - V_D)/e$, where $C_g$ is the capacitance per unit area of the gate dielectric with ~19.5 μF/cm² estimated through Hall measurements in Fig. 3(c),[25] $V_g$ is the applied gate voltage, and $V_D$ is a voltage corresponding to the Dirac point. $V_D$ is estimated to be -1.54 V. The crossover of the Dirac point is not shown in Fig. 2(b), since our measurement is limited by the electrochemical reaction between the ion gel and TI channel beyond the gate bias of -1.5 V.

Ideally, the value of resistance is to be infinity at the Dirac point. However, according to a previous theoretical prediction,[26] the inhomogeneous puddle-dominated impurities provide the spatially fluctuating local conductivities and contributes to the finite resistivity near the Dirac point, which can be calculated more accurately by the effective medium theory (EMT). As the EMT is above the scope of the present study, the modulation of the resistance in Fig. 2(b) has been fitted by the Boltzmann transport theory using the effective interaction parameter $r_s = e^2 / \kappa \hbar v_F$ and a characteristic density $n_0 = (m^* v_F)^2 / (4\pi \hbar^2)$, where $v_F$ is the Fermi velocity and $\kappa$ is approximately half of the dielectric constant of $Bi_2Se_3$. The conductivity can be calculated by defining[26]

$$\sigma_B = \frac{1}{8} \frac{e^2}{h} \frac{n}{n_{imp}} \frac{1}{F[\eta r_s / 2]} \quad , \quad \frac{F[x]}{x^2} = \frac{\pi}{4} + 3x - \frac{3x^2 \pi}{2} + x(3x^2 - 2) \frac{\arccos[1/x]}{\sqrt{x^2 - 1}} \quad ,$$

$$\eta[n/n_0] = \frac{\sqrt{n_0}}{\sqrt{n_0} + \text{sgn}(n)\sqrt{|n|}} ,$$

where $n_{imp}$ is the charged impurity density, and sgn is the signum function. From the fit in Fig. 2(b), we estimate $r_s$ of 0.9, $n_0$ of 5 × 10¹⁴/cm², and $n_{imp}$ of 5 × 10¹³/cm². A high value of $n_{imp}$ indicates the charged impurities play an important role for the electronic transport, leading to a finite saturating value of resistance as the gate bias moves the Fermi level toward the Dirac point ($V_D$ = -1.54 V). The charged impurities at low carrier density close to the Dirac point provide inhomogeneous electron-hole puddles, where the conductivity will be approximately a constant over a finite range of external gate voltage, providing an explanation for the observed resistivity plateau in Fig. 2(b).[27-29] The Hall measurements without and with the ionic gate indicate that the carrier density of the TI channel increases with ionic gating by a factor of ~2 as can be compared in Fig. 1(c) and 3(f). As shown in the fit of Fig. 2(b), the Boltzmann transport model cannot accurately explain the data around the Dirac point as it gives zero conductance (infinite conductance) at the Dirac point.

The output curves of an ionic gel gated TI FET for 5 different gate voltages ($V_g$) show a linear increase with a small hysteresis in the $V_{sd}$ sweeping range of -1.2 to 1.2 V (not shown). The transconductance ($g_m = \partial I_{sd}/\partial V_g$) at $V_{sd}$ = -1.2 V and field effect mobility derived from the transfer curve is shown in Fig. 3(a) and Fig. 3(b) respectively. Using a formula of $\mu_{FE} = 1/C_g \times d\sigma_{2D}/dV_g$,[30] where $\sigma_{2D}$ is the conductivity of the channel with a channel length of 100 µm and a channel width of 20 µm, the µ$_{FE}$ is estimated to increase from 68 to 194 cm$^2$/V·s by applying gate bias from 0.5 to -1.1 V. We perform the Hall measurements and observe a variation of the channel conductivity as a function of gate voltage at room temperature. The carrier density and mobility at 300 K is plotted in Fig. 3(c). The carrier concentration of the TI channel decreases as the gate bias decreases from 1 to -1 V, while the mobility increases. When the negative gate bias is applied, the Fermi level is expected to shift toward the Dirac point. Consequently, reduced number of electrons in the transport may cause an increase of the channel resistance, and the mobility increases as the carrier density decreases.[31,32] Unlike two-dimensional electron layers in semiconductors, where the charge carriers become immobile at low densities, the carrier mobility in graphene can remain high, even when their density vanishes at the Dirac point.[33] For monolayer graphene, the mobility, limited by short range scattering, was found to be inversely proportional to the carrier density.[32] This inverse relationship is also observed from the Hall measurements as shown in Fig. 3(d) which indicates that the carrier transport is also affected by the surface states. Note that the carrier densities from the Hall measurements in Fig. 3(d) are very similar to the ones extracted from the transfer curves in Fig. 3(b). As shown in Fig. 3(c), the variation of the carrier density is ~3.8 times, which is larger than the variation of mobility (~2.3 times). Therefore, the observed gate dependent modulation of the channel conductance can be attributed to carrier modulation in the TI channel.

In Fig. 3(e), the conductance $\sigma = e n_{2D} \mu_H$ as a function of the carrier density is fitted by a recent theoretical study, which predicts the conductivity as being limited by charged impurity scattering assuming a linear Dirac band,[27,34] $\sigma(n) \sim C |n/n_{imp}| [e^2/h]$, where $C$ is a fitting constant ($C = 20$).[28] From the fits in Fig. 3(e), we identify the impurity density ($n_{imp}$) ranges from $4.28 \times 10^{13}/cm^2$ for $n_{2D} < 12 \times 10^{13}/cm^2$ to $2.6 \times 10^{14}/cm^2$ for $n_{2D} > 12 \times 10^{13}/cm^2$, which is very close to the value estimated from the Boltzmann transport theory in Fig. 2(b). On the other hand, as seen in Fig. 3(f), the variation of the carrier concentration is comparably smaller, as temperature drops from 300 to 2 K, than that of Hall mobility. Therefore, the temperature dependent conductance modulation in Fig. 1(b) can be attributed to a change in the mobility, rather than to carrier modulation. The fact that the background carrier densities do not change much with temperature indicates full ionization of electron donors in the TI even at low temperatures.[35] The Hall mobility increases by ~2.23 times, as temperature drops from 300 to 30 K, indicating photon scattering process.[35] An almost constant mobility below ~30 K suggests static disorders induced by the Se vacancy or penetration of [TFSI]$^-$ as the dominant scattering mechanism.[15,36,37]

In summary, we have reported a high modulation ratio of the resistance from $Bi_2Se_3$ TI FETs with an ionic liquid top gate at room temperature. With a gate voltage of -1 V, the device presents a resistance modulation ratio of 365%. According to the Hall measurements, the majority carriers in the TI channel are electrons, and the carrier density of the channel effectively tuned by an ionic liquid gate is responsible for conductance modulation. On the other hand, the temperature dependence of conductance modulation is due to an increase of the mobility at low temperatures. The observed large conductance modulation with ion gel gating presents an interesting route towards the applications of topological insulator based devices at room temperature.

This work is partially supported by the Singapore Ministry of Education Academic Research Fund Tier 1 (R-263-000-A75-750) and by NSF DMR-0845464 and ONR N000141210456 at Rutgers.

Figure captions

FIG. 1. (a) Optical transmission of $Bi_2Se_3$ films in the UV-Vis-IR range. (b) Temperature dependent resistance of a 20 QL $Bi_2Se_3$ film. (c) Hall measurement of a TI sample at 2 K. (d) $I_{sd}$ - $V_{xx}$ curves at 2 and 300 K. The samples are measured without an ionic gate using a four probe technique.

FIG. 2. (a) Optical micrograph of the patterned Hall bar with an ion gel gate and measurement diagram of the FET device. (b) Resistance versus gate voltage with different sweep ranges along with a Boltzmann transport theory fit.

FIG. 3. The transconductance (a) and field effect mobility (b) as a function of gate voltage. (c) The carrier density and Hall mobility as a function of gate bias at 300 K. (d) Hall mobility as a function of carrier density at 300 K. (e) Conductance as a function of the carrier concentration with a linear fitting. (f) Temperature dependance of the carrier density and mobility without gate bias.

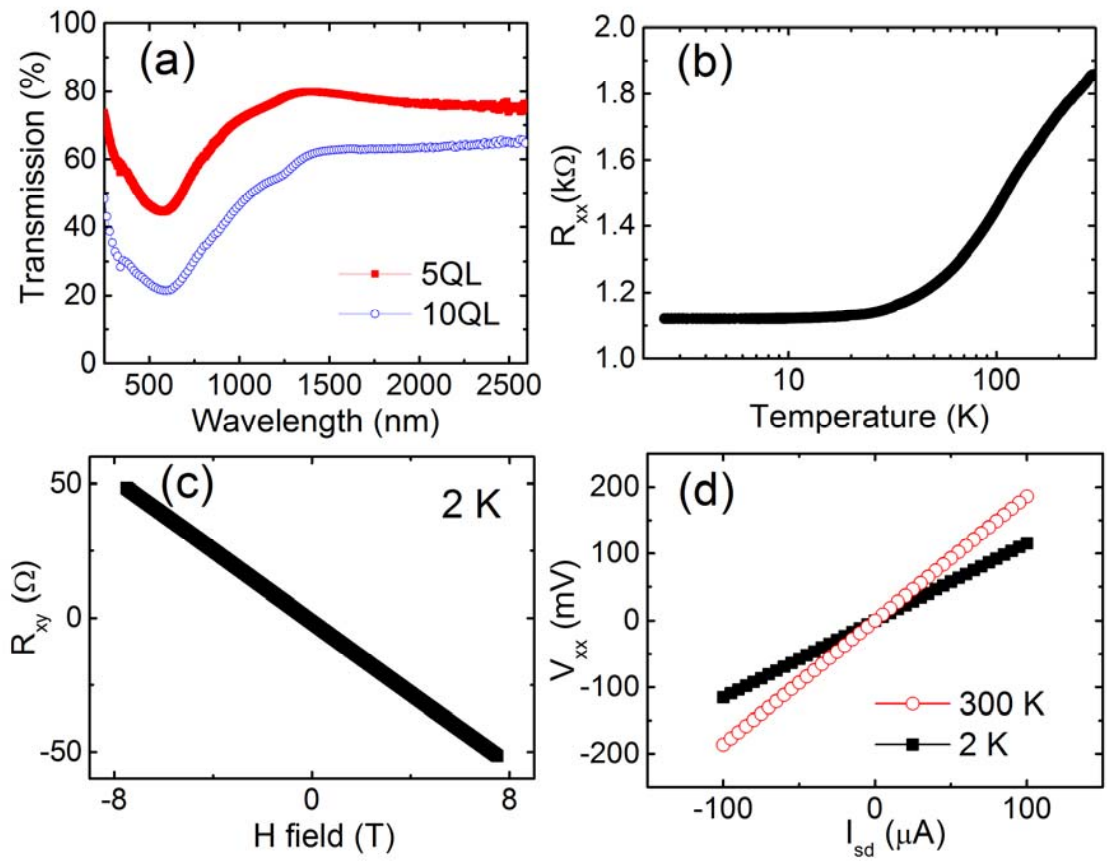

Figure 1

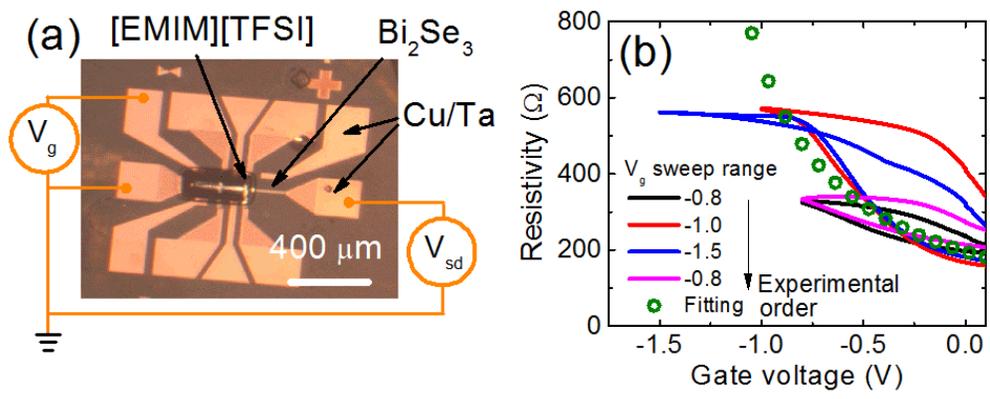

Figure 2

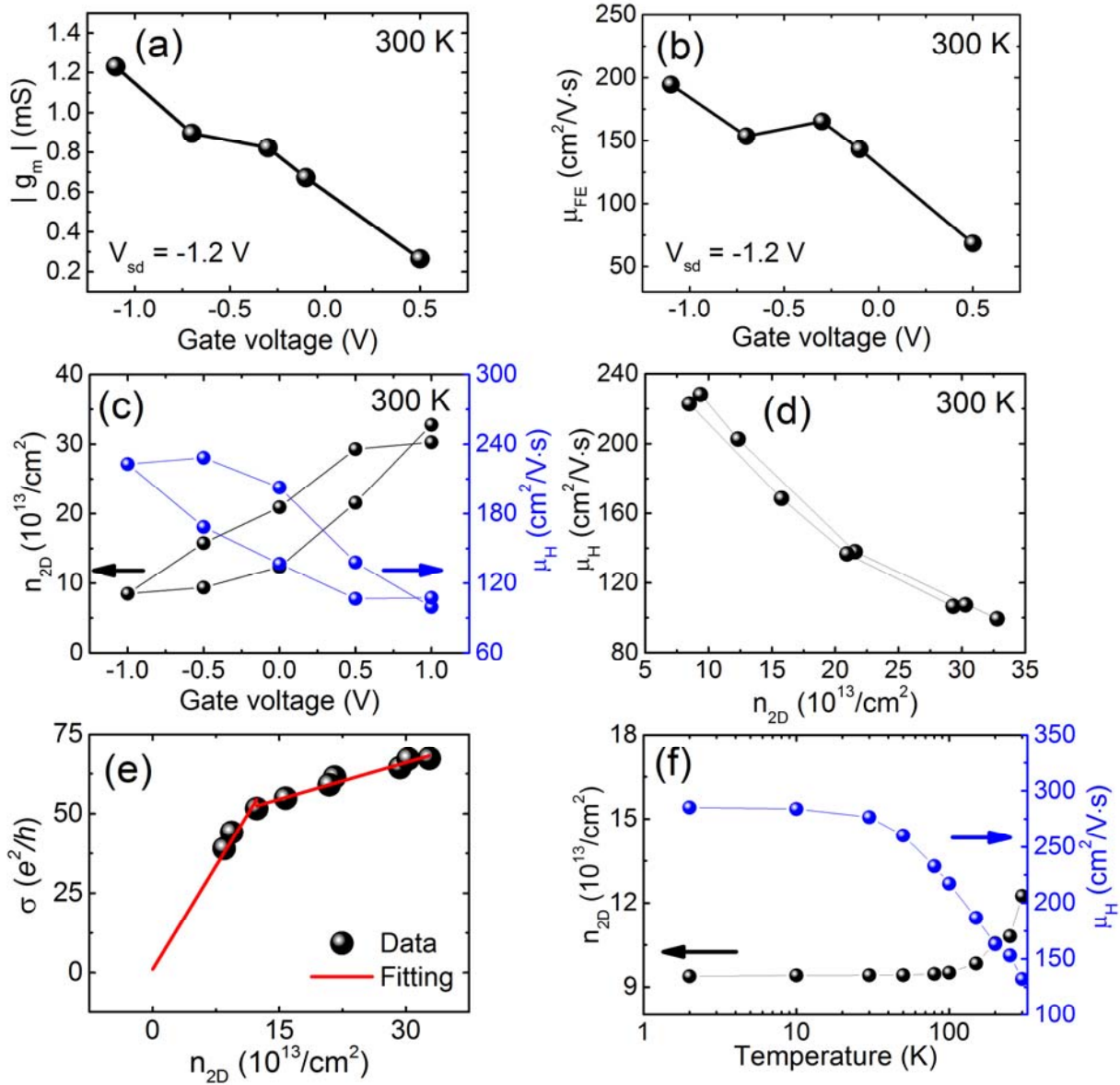

Figure 3